\documentclass[preprint,showpacs,preprintnumbers,amsmath,amssymb]{revtex4}

\usepackage{graphicx}
\usepackage{dcolumn}
\usepackage{bm}

\begin{document}


\title{Force-velocity correlations in a dense, collisional, granular flow}

\author{Emily Gardel}
\author{Ellen Keene}
\author{Sonia Dragulin}
\author{Nalini Easwar}%
\affiliation{Department of Physics, Smith College, Northampton,
U.S.A.}%

\author{Narayanan Menon}%
\affiliation{%
Department of Physics, University of Massachusetts, Amherst,U.S.A.
}%

\date{April 15, 2002}

\date{\today}

\begin{abstract}
We  report measurements in a 2-dimensional, gravity-driven,
collisional, granular flow of the normal force delivered to the wall
and of particle velocity at several points in the flow. The wall
force and the flow velocity are negatively correlated. This
correlation falls off only slowly with distance transverse to the
flow, but dies away on the scale of a few particle diameters
upstream or downstream. The data support a picture of short-lived
chains of frequently colliding particles that extend transverse to
the flow direction, making transient load-bearing bridges that cause
bulk fluctuations in the flow velocity.  The time-dependence of
these spatial correlation functions indicate that while the
force-bearing structures are local in space, their influence extends
far upstream in the flow, albeit with a time-lag.  This leads to
correlated velocity fluctuations, whose spatial range increases as
the jamming threshold is approached.
\end{abstract}

\maketitle

Sand flowing down a long, vertical, pipe does not accelerate in the
direction of gravity because the walls of the pipe support the
weight of the sand.  When the outlet of the pipe is constricted, the
flow slows down, but also becomes variable in time, with large
regions of material appearing to move in unison. Momentum balance
dictates that if the flow speed fluctuates in time, then the forces
supplied by the walls must also vary in time.  The average force
borne by the walls of the pipe does not significantly change with
the average flow velocity, therefore all dynamical information
regarding the state of the flow must be contained in these force
fluctuations.

In this article, we report measurements of the spatial and temporal
correlations of fluctuations of the flow velocity and of the wall
forces as the flow gets slower.  We find that short-lived, local
fluctuations in the wall force produce fluctuations of flow velocity
at hydrodynamic time scales and over large length scales, which
increase as jamming is approached.  We also address the closely
related issue of the spatial organization of these forces. When the
flow is permanently arrested, the weight of the grains is
communicated to the walls by a spatially heterogeneous web of forces
called force chains \cite{Dantu&Travers,Liu1995,Howell1999}. The
force-velocity correlations we measure are consistent with a picture
in which close to jamming, the instantaneous stress configuration is
similar to that seen in the static case, except with short-lived
force chains that temporarily hold up the flow, then disintegrate
and allow the flow to accelerate again. Remarkably, we observe these
signatures of dynamic force chains in a flow where stresses are
communicated chiefly by collisions, with grains remaining in contact
for only a small fraction of the time that they are in flight
between collisions.

\begin{figure}
\includegraphics[width=.5\textwidth]{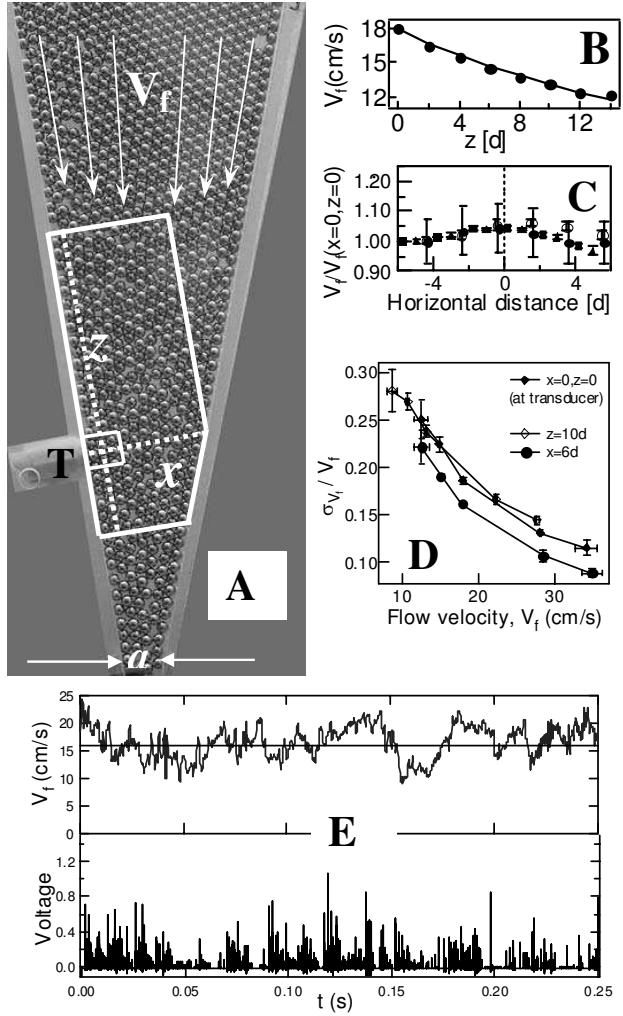}
\caption{\label{sketch} (A) Image of the lower part of the hopper.
The flow rate is controlled by varying the outlet $a$. The force
transducer is marked 'T'.The large white box indicates the field of
view ($\approx 10d$ x $20d$) used to visualize the flow field at
$500\thinspace fps$. We measure fast fluctuations in the flow
velocity $V_{f}$ at a higher rate of $4000\thinspace fps$ in a field
of view whose size is shown by the small square ($2.5d$ x $2.5d$).
(B) $V_{f}$ vs. position $z$ going up the wall from the transducer,
for $a=4d$. The solid line is the $z$-dependence expected from the
continuity equation for an incompressible flow. (C) $V_{f}$
normalized to the velocity at the transducer vs. position going
horizontally across from the transducer, for several flow rates. The
dashed line is the midpoint of the flow. (D) The normalized standard
deviation in flow speed, $\sigma_{V_{f}}/V_{f}$, as a function of
$V_{f}$ measured at three locations: at the transducer, $6d$ across
from it, and $10d$ up along the wall. (E) Time-dependence of the
transducer voltage (proportional to force) and $V_{f}$ for $a=4d$.}
\end{figure}

Most previous experimental investigations \cite{Majumdar05} of force
chains in granular flows have been in the quasi-static regime, where
grains are essentially in continuous contact.  However, some
simulations of collisional flows show evidence of large-scale
 stress-bearing structures. Simulations of pipe flows have
observed long-range decay of stress in specific directions
\cite{Denniston1999} as well as of linear structures made up of
frequently colliding particles that transport stress efficiently
\cite{Ferguson2004}. In studying the initiation of flow by opening
the bottom of a container, it was found \cite{LudingDuran96} that
the flow started with chunks of material falling out of the
container; structures labelled "dynamic arches" held up the rest of
the material temporarily, until they broke under the remaining load.
A broader understanding of such intermediate-scale stress-bearing
structures would  help identify the microscopic origins of the
non-local rheology reported in a variety of flows
\cite{GDRMidi2004,Pouliquen2004, Ertas2002}.

We study the flow of smooth, spherical steel balls of diameter
$d=3.125 \thinspace mm$ contained in a 2-dimensional hopper. The
flow velocity is controlled by varying the width, $a$, of the outlet
from $3d$ to $8d$, with the sides held at a fixed angle of $10^\circ
$ to the vertical. The flow field inside the hopper is measured by
imaging at a rate of 500 frames/second ($fps$) the region shown
inside the white box in Fig \ref{sketch}A. In Fig \ref{sketch}B, we
show the variation of the time-averaged flow velocity, $V_{f}$, as a
function of position in the hopper. In the vertical direction,
$V_{f}$ varies inversely as the width of the hopper; in the
horizontal direction (Fig \ref{sketch}C) the velocity profile shows
only a weak spatial dependence with a large slip at the wall, and a
maximum at the centre of the channel.  The strain rate at the
transducer, $\partial V_{f}(x)/\partial z\approx1.5 sec^{-1}$, is
nearly independent of the flow rate. Fluctuations about this average
velocity profile are anisotropic, with bigger fluctuations along the
flow; this is agreement with measurements
\cite{Choi2004,MokaNott2005} at the walls of 3-dimensional dense
flows. As the flow velocity is reduced, these fluctuations increase
as shown in Fig \ref{sketch}D, where we plot the normalized standard
deviation of the flow velocity, $\sigma_{V_{f}}/V_{f}$, against the
local flow velocity, $V_{f}$, for three different locations in the
cell. At all three locations, fluctuations grow strongly relative to
the mean as jamming is approached.

As argued previously, the dynamical origin of these velocity
fluctuations must lie in fluctuations in the forces exerted by the
walls. We measure this force by means of a normal force transducer
embedded in the side wall; the active surface of the transducer is
large enough to accommodate exactly one ball. The voltage output of
the transducer (Fig. \ref{sketch}E), sampled at $100 \thinspace kHz$
is converted into a force by calibration with impacts and static
loads of known magnitude. Over the entire range of the flow
velocities explored in this article, the forces exerted against the
wall are primarily collisional rather than frictional, with balls
making repeated but isolated impacts against the transducer. It is
not possible to discern the collisional nature of the forces from
inspection of video images since the inter-particle separations can
be extremely small at these high packing fractions. Simulations
\cite{Campbell2002} show that small changes in packing fraction can
drive a change in the microscopic mechanism of momentum transfer
from collisions to friction. We have previously studied
\cite{Longhi02} the statistics of the impulsive forces as a function
of the flow rate, finding that the distribution of forces remained
broad at all flow rates, with an exponential tail, just as in static
granular media. In order to study the correlation of the flow and
the forces, we make measurements of the velocity at $4000 fps$,
synchronized carefully with the force measurement(Fig.
\ref{sketch}E). To achieve these higher imaging speeds without
sacrificing spatial resolution, we are obliged to make the velocity
measurement in a smaller window as indicated in \ref{sketch}A;
measurements are made for several positions of this window: along
the wall upstream ($+z$ direction) and downstream ($-z$) of the
transducer, as well as normal to the wall ($+x$).

\begin{figure}
\includegraphics[width=.5\textwidth]{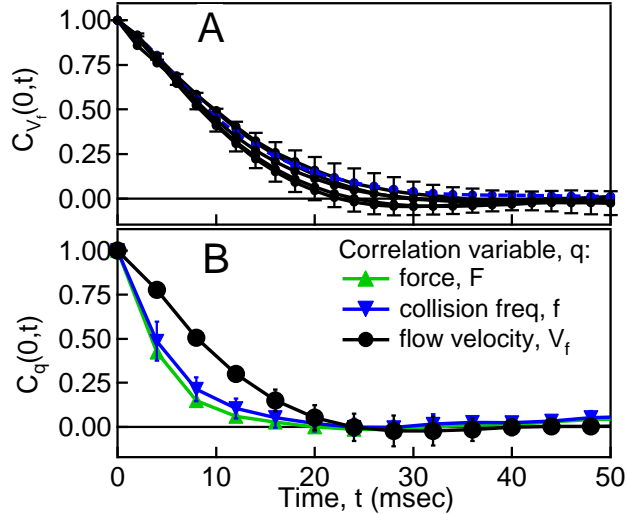}
\caption{\label{autocorr} (A) Autocorrelation function,
$C_{V_{f}}(0,t)$ of the flow velocity, $V_{f}$ at the transducer
$(z=0,x=0)$ as a function of time, $t(msec)$ for openings $a/d=3.04,
3.28, 4, 6, 8$. We also include data for $a/d=3.04$ at $z=10d$. The
decay time of the autocorrelation does not change as $V_{f}$ changes
by a factor $4.25$.(B) The autocorrelation functions of collision
frequency, $f$, and average force, $F$, decay faster than that of
$V_{f}$. The error bars are standard deviations between 5 data sets,
each consisting of 40,000 video frames acquired at $4000 fps$.}
\end{figure}

The temporal characteristics of the fluctuations in flow velocity
$V_{f}$, are quantified in Fig. \ref{autocorr}A, where we show the
autocorrelation function $C_{V_{f}}(0,t)$ \cite{notation} of the
velocity at the transducer for several flow speeds. Unlike the
magnitude of the fluctuations in $V_{f}$, the decay time is not
sensitive to the distance from the jamming threshold:  within
statistics it is unchanged over a velocity range of $8$ to $34
\thinspace cm/s$.  (This differs from Ref. \cite{Choi2004} where
mean-squared particle displacements collapse when plotted against
$V_{f}t$, the distance the particle is advected; presumably the
difference is that we are in an inertial regime whereas they are in
a quasistatic regime where geometry is the dominant consideration).
The fluctuations in forces decay faster than the fluctuations in
flow velocity, as can be seen in Fig. \ref{autocorr}B, where we
compare the autocorrelation of $V_{f}$, of force, $F$, and of the
collision frequency, $f$.


\begin{figure}
\includegraphics[width=.5\textwidth]{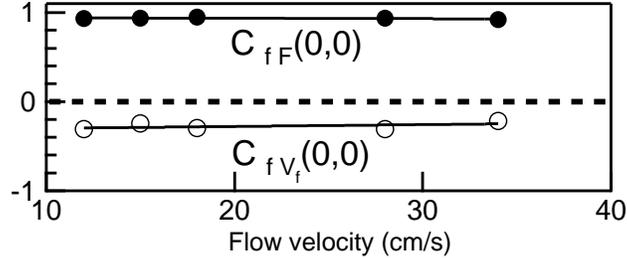}
\caption{\label{C_00} Equal-time cross correlations $C_{fF}(0,0)$
and $C_{f{V_{f}}}(0,0)$ at the transducer (x=0, z=0) as a function
of average flow velocity $V_{f}$, showing the high positive
correlation of $F$ and $f$, and the negative correlation between $f$
and $V_{f}$ at all flow velocities.}
\end{figure}

Even though the collision frequency $f$ and
 velocity, $V_{f}$ fluctuate on different time scales, their equal-time
 cross-correlation, $C_{f{V_{f}}}(0,0)$, shown in Fig.\ref{C_00}, establishes
 that these quantities are anti-correlated: higher-than-average
 collision frequencies are accompanied by negative
 fluctuations in velocity. The same is true of the correlation between
 velocity and average force: large forces accompany negative velocity
 fluctuations. This is implied by the strong positive correlation shown in Fig.
 \ref{C_00} of $f$  and $F$ at all flow velocities. This is not a surprising
 correlation, after all, frequent collisions against the wall
 generally would indicate greater momentum transfer to the wall,
 however, the degree of the correlation is very strong. Thus most
 of the information in force fluctuations is carried by the
 frequency of collisions, and very little by fluctuations in the
 magnitude of the impulses.

\begin{figure}
\includegraphics[width=.5\textwidth]{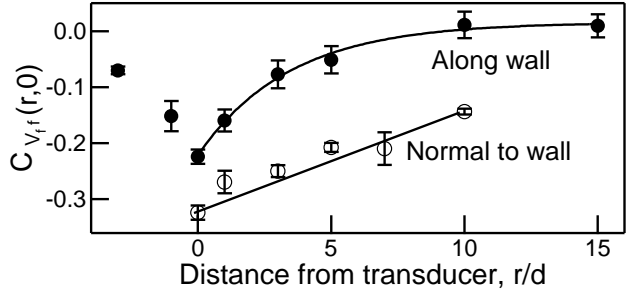}
\caption{\label{spacecorr} Equal-time cross-correlation
 between the collision frequency $f$ at the
transducer, and velocity fluctuations at different locations along
the wall $C_{f{V_{f}}}(z,0)$, and normal to the wall
$C_{f{V_{f}}}(x,0)$ (this trace is vertically offset for clarity).}
\end{figure}

Is the force at the wall anticorrelated only with the velocity
exactly at that point? We investigate this question by plotting in
Fig. \ref{spacecorr}, the equal-time cross-correlation between the
collision frequency $f$ at the transducer, and velocity fluctuations
at several locations along the wall $C_{f{V_{f}}}(z,0)$, and at
other locations normal to the wall $C_{f{V_{f}}}(x,0)$.  The
greatest anticorrelation is at $x=z=0$, that is, between forces
measured at the transducer and $V_{f}$ at that location. These
correlations decay anisotropically from this point. Along the wall,
the correlation dies off rapidly as a function of $z$, with a decay
length of $2$ to $3\thinspace d$. Into the flow, though, the
correlation decays approximately linearly in $x$, going to zero only
at the far wall. Thus the equal-time correlation is consistent with
a high-collision-rate structure that is only a few beads wide in the
flow direction, but can span the entire channel transverse to the
flow. This can be identified with the collision chains seen in the
simulations of \cite{Ferguson2004}. Similar structures have also
been visualized in the slow flow of soft photoelastic discs in a 2D
hopper similar to ours \cite{Vivanco}.


\begin{figure}
\includegraphics[width=.5\textwidth]{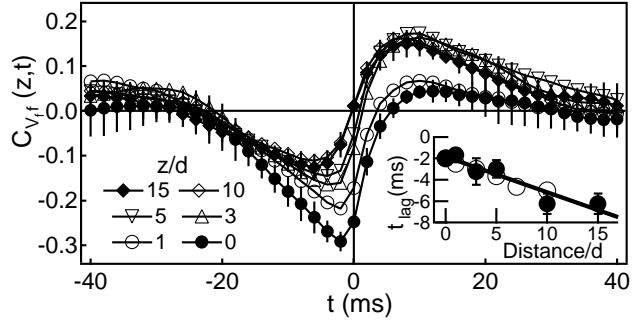}
\caption{\label{vf_corr} The time-dependent cross correlation
$C_{f{V_{f}}}(z,t)$ against delay time, $t$ for different $z$ along
the wall. The maximum of the negative correlation occurs at a
negative delay time $t_{lag}$. Inset: $t_{lag}$ versus distance,
$z/d$ ($\bullet$) and $x/d$ ($\circ$).}
\end{figure}

Even though the equal-time correlations shown in Fig.
\ref{spacecorr} die off quickly in the upstream direction, the
high-collision rate events can have an effect far upstream. This is
most clearly revealed in the time-dependent cross correlation
$C_{f{V_{f}}}(z,t)$ shown in Fig. \ref{vf_corr}.  We first focus on
the data ($\bullet$) for $z=0$ .  It is apparent that the greatest
anticorrelation is at negative times and not at t=0. Thus, the
collision rate builds up and is maximum at a time {$t_{lag}$} before
the flow slows down. There is also a broad peak for positive times,
indicating that a period of acceleration is likely to follow, and
that there is a suppressed probability for the formation of another
collisional "arch" immediately after. The other curves shown in Fig.
\ref{spacecorr} represent $C_{f{V_{f}}}(z,t)$ for increasing values
of $z$, i.e. for positions along the wall, upstream from the
transducer. The anticorrelation at the minimum of the curve
$(t=t_{lag})$ is larger, and diminishes much slower with distance,
than does the anticorrelation at $t=0$. Thus even though the
equal-time correlation indicates a thin, local structure, the effect
on the flow is extremely long-range.  $t_{lag}(z)$, the time
required for the information propagate a distance $z$ upstream, is
shown in the inset.  The slope represents the velocity at which the
force information propagates upstream; this velocity is more than an
order of magnitude greater than the flow speed, and is comparable to
$d \langle f\rangle$, the particle diameter multiplied by the mean
collision frequency. Thus, the communication is very coherent, akin
to a longitudinal sound mode.


\begin{figure}
\includegraphics[width=.5\textwidth]{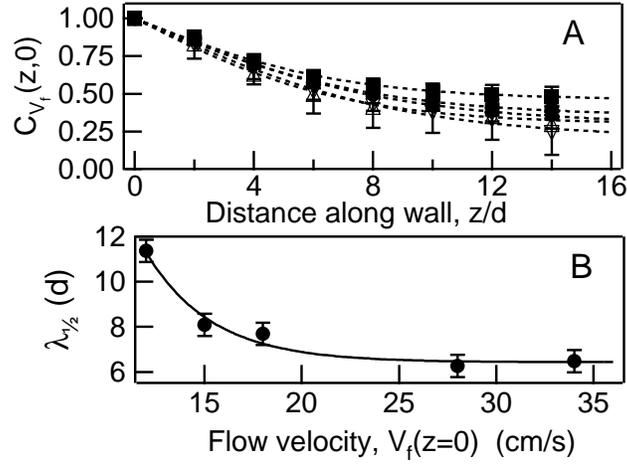}
\caption{\label{lengthscale} (A) Equal-time spatial cross
correlations $C_{V_{f}}(z,0)$  between velocities $V_{f}$ at
different $z$  along the wall. (B) The distance, $\lambda_{1/2}$, at
which the spatial correlation decays to 1/2, as a function of flow
velocity, $V_{f}$.}
\end{figure}
This long-range effect of clogging by transient arches leads to a
flow with pronounced spatial correlations: the extreme version of
this occurs when a stable arch forms across the channel - the flow
comes to a halt everywhere in the hopper. However, there is a
continuous approach to this limit as the flow velocity is slowed
down towards jamming. This is shown in Fig. \ref{lengthscale}A where
we plot for several flow speeds, the spatial correlation of the
velocity field, $C_{V_{f}}(z,0)$, as a function of position along
the wall. The correlation decays slower in space as the flow slows
down. This is demonstrated in Fig. \ref{lengthscale}B where we show
the distance, $\lambda_{1/2}$, at which the correlation falls to
$1/2$.

 Thus the approach to jamming is
characterized by both an increasing correlation length scale in the
velocity fluctuations (Fig. \ref{lengthscale}) as well as an
increasing amplitude (Fig. \ref{sketch}D) for the fluctuations. We
have also previously observed in this geometry \cite{Longhi02} that
a time-scale diverges: the collision time distribution goes to a
power-law with an exponent of $-3/2$. The driving force for the
slow-down and arrest of the flow appears to be collisionally
stabilized structures reminiscent of the force chains observed in
static granular packings. In our experiments, these collision chains
are not directly observed but are identified via their
anticorrelation with the velocity field: a high collision rate
immediately precedes the slowing down of the flow. At any instant,
the correlations indicate that these transient structures are thin
in the flow direction and long-ranged transverse to the flow. The
effect of these localized structures is propagated rapidly to
upstream parts of the flow, signalling a temporary clogging of the
flow downstream. We are currently exploring similar phenomena in 3D
flows \cite{3Dflow}, where force measurements reveal both
collisional and frictional regimes.


We thank F. Rouyer for her help with the experiment, and B.
Chakraborty, A. Ferguson, S. Tewari, J.W. Landry, A.D. Dinsmore for
useful discussions. We acknowledge support from NSF-DMR 0305396, and
NSF-MRSEC 0213695.



\begin{thebibliography}{14}


\bibitem{Dantu&Travers}
P. Dantu, Ann. Pont Chaussees IV, 144 (1967); T. Travers et al.,
Europhys. Lett. {\bf 4}, 329 (1987).

\bibitem{Liu1995}
C. H. Liu et al., Science {\bf 269}, 513 (1995).

\bibitem{Howell1999}
D. Howell, R. P. Behringer, C. Veje, Phys. Rev. Lett. {\bf 82}, 5241
(1999).

\bibitem{Majumdar05}
T. Majumdar, R.P. Behringer, Nature, (2005).

\bibitem{Denniston1999}
C. Denniston, H. Li, Phys. Rev. E {\bf 59}, R3289 (1999).

\bibitem{Ferguson2004}
A. Ferguson, B. Fisher and B. Chakraborty, Europhys. Lett.
\textbf{66}, 277 (2004).

\bibitem{LudingDuran96}
S. Luding,  J. Duran, E. Clement, J. Rajchenbach, J. Phys. I
\textbf{6}, 823 (1996); J. Duran et al., Phys. Rev. E \textbf{53}
1923 (1996).

\bibitem{GDRMidi2004}
GDRMidi, Eur. Phys. J. E \textbf{14}, 34 (2004)

\bibitem{Ertas2002}
D. Ertas, T.C. Halsey, Europhys. Lett. \textbf{60}, 931 (2002).

\bibitem{Pouliquen2004}
O. Pouliquen, Phys. Rev. Lett. \textbf{93}, 248001 (2004).

\bibitem{Choi2004}
J. Choi, A. Kudrolli, R.R. Rosales, M. Bazant, Phys. Rev. Lett.
\textbf{92}, 174301 (2004).

\bibitem{MokaNott2005}
S. Moka, P. Nott, Phys. Rev. Lett. \textbf{95}, 068003 (2005).

\bibitem{Campbell2002}
C.S. Campbell, J. Fluid Mech., \textbf{465}, 261 (2002).

\bibitem{Longhi02}
E. Longhi, N. Easwar, N. Menon, Phys. Rev. Lett. \textbf{89}, 045501
(2002).

\bibitem{notation}
We denote the time- and space-correlation of variables $q$ and $p$
by $C_{qp}(r,t)$ $\equiv$ $\langle \tilde{q}(0,t_{o})\tilde{p}(r,
t_{o}+t)\rangle$ where $\tilde{q(t)}= q(t)-\langle q\rangle$. The
angle brackets, $\langle\rangle$, denote an average over the initial
time, ${t_{o}}$; since the system is not homogeneous in space, no
average is taken over the initial position, typically chosen to be
the location of the transducer. The subscript $q$ is not repeated
for autocorrelation functions.

\bibitem{Vivanco}
F. Vivanco, F. Melo, S. Rica, private communication.

\bibitem{3Dflow}
E. Seitaridou, E. Gardel, N. Easwar, N. Menon, in preparation.





\end{thebibliography}

\end{document}